\documentclass[twocolumn,superscriptaddress,aps,prb,floatfix]{revtex4-1}

\usepackage{graphicx}
\usepackage{dcolumn}
\usepackage{bm}
\usepackage{color}
\usepackage[caption=false]{subfig} 

\usepackage{listings}

\definecolor{dkgreen}{rgb}{0,0.6,0}
\definecolor{gray}{rgb}{0.5,0.5,0.5}
\definecolor{mauve}{rgb}{0.58,0,0.82}

\usepackage{amsthm}
\usepackage{amsmath}
\usepackage{amssymb}

\newcommand{\Htc}{H_{\mathrm{TC}}}
\newcommand{\figref}[1]{Fig. \ref{#1}}

\newtheorem{theorem}{Theorem}[section]

\begin{document}


\title{Engineering autonomous error correction in stabilizer codes at finite temperature}

\author{C. Daniel Freeman}
\email{daniel.freeman@berkeley.edu}
\affiliation{Berkeley Quantum Information \& Computation Center, University of California, Berkeley, CA 94720, USA}
\affiliation{Department of Physics, University of California, Berkeley, CA 94720, USA}

\author{C. M. Herdman}
\affiliation{Institute for Quantum Computing, University of Waterloo, Waterloo, ON N2L 3G1, Canada}
\affiliation{Department of Chemistry, University of Waterloo, Waterloo, ON N2L 3G1, Canada}
\affiliation{Department of Physics \& Astronomy, University of Waterloo, Waterloo, ON N2L 3G1, Canada}

\author{K. B. Whaley}
\affiliation{Berkeley Quantum Information \& Computation Center, University of California, Berkeley, CA 94720, USA}
\affiliation{Department of Chemistry, University of California, Berkeley, CA 94720, USA}

\date{\today}

\begin{abstract}
 We present an error correcting protocol that enhances the lifetime of stabilizer code based qubits which are susceptible to the creation of pairs of localized defects (due to string-like error operators) at finite temperature, such as the toric code. The primary tool employed is dynamic application of a local, unitary operator which exchanges defects and thereby translates localized excitations.  Crucially, the protocol does not require any measurements of stabilizer operators, and therefore can be used to enhance the lifetime of a qubit in the absence of such experimental resources.
\end{abstract}

\maketitle

\section{Introduction}
\label{sec:Intro}

For the past two decades, significant effort has gone into devising schemes for encoding quantum information in reliable and retrievable forms.  Stabilizer error correcting codes are thought to be an effective strategy for performing this encoding, because they allow an efficient means of detecting and correcting errors. Among these, topological stabilizer codes (or topological quantum memories), are particularly promising strategies for storing quantum information due to their intrinsic robustness to errors at zero temperature, their ability to be efficiently implemented via a local Hamiltonian\cite{Fowler2012}, as well as the existence of efficient strategies for performing error detection and correction\cite{Wang2009,Duclos-Cianci2010,Fujii2014} which have been demonstrated in recent experiments\cite{Nigg2014}.  Several exhaustive studies have been performed on calculating error thresholds for these topological codes, like Kitaev's toric code, both in the presence and absence of error correcting protocols\cite{Bravyi1998,Wang2009,Fowler2009,Duclos-Cianci2010,Wootton2012}.

However, these topological codes are well known to be poor \emph{passive} quantum memories at finite temperature in less than four spatial dimensions\cite{Nussinov2008a,Nussinov2009a,Castelnovo2007a,Alicki2009a,Chesi2010a,Chesi2010b, Yoshida2011b, Viyuela2012,Watson2014,Freeman2014} (for a thorough review, see Ref.~\onlinecite{Brown2014}).  For physically realistic coupling to an environment, local noise processes drive the creation of localized defects.  In the absence of an error correcting protocol, the propagation of these defects can then lead to decoherence of the memory.  For the case of the toric code, these error strings are particularly pathological, and cause the maximum lifetime of an encoded qubit to decay exponentially with temperature with a timescale independent of system size\cite{Alicki2009a}.  While experimentally intractable, fault tolerant topological quantum memories are known to exist in four and six dimensions\cite{Dennis2002,Bombin2013}.

On the other hand, a variety of active error correction protocols exist for efficient detection and correction of errors.  As long as error rates and the temperature are low or, alternatively, as long as detection and correction are fast enough, the lifetime of these codes can in principle be extended indefinitely.  But these decoding strategies implicitly rely on resources that may not always be available or efficiently physically implementable.  For example, performing a measurement on a quantum system requires a fresh ancilla qubit for each measurement.  Thus, continuously measuring any quantum system requires continuously recycling ancilla qubits for measurement---a procedure which will necessarily be rate limiting for near term quantum architectures\cite{Wecker2014}. 

An error correcting strategy for topological codes \emph{without} the need for stabilizer measurement is desirable.  At face, ignoring the power of the stabilizer group will assuredly provide a suboptimal strategy.  But given limited resources and rates of measurement it is worthwhile to understand the limits of strategies which do not require syndrome measurements, and to determine if such strategies can augment known decoding schemes.  

We provide here a new protocol for error correction of pairs of localized defects which modifies an existing dissipative protocol.  We achieve this by applying a specially designed sequence of unitary operators to a code of choice.  This pattern of operators is designed to encourage defects in the system to dissipate more quickly.  In this work, we explicitly treat the theory for the 1D Ising model at finite temperature, and describe how this approach may be extended to other stabilizer codes, such as the toric code.  While dissipative protocols have previously been employed to generate hamiltonians\cite{Herdman2010, Weimer2010}, to prepare encoded ground states\cite{Dengis2014}, to mediate long range interactions\cite{Hutter2012a, Pedrocchi2013, Kapit2015}, and to ``trap'' defects\cite{Bardyn2015,Fujii2014}, a dissipative protocol that explicitly targets string-like error processes has not been proposed to date.  While it does not completely eliminate errors, the protocol presented here provides a significant enhancement of the lifetime of a finite-size system.

It is known that stabilizer Hamiltonians at finite temperature in dimension less than three have a system-size independent upper bound to their lifetime\cite{Bravyi2009, Alicki2009a, Chesi2010b, Yoshida2011b, Hastings2011a, Landon-Cardinal2013, Temme2014, temme2015fast}.  These ``no-go'' theorems necessarily limit the extent to which the method proposed here can be carried out.  In fact, a size-independent constant enhancement of a system's lifetime may be the best one can get with a purely local unitary protocol like the one presented here.  Thus, this scheme, by itself, will not generate a topologically protected quantum memory at finite temperature for one or two dimensions. It is nonetheless worthwhile to understand how far purely local protocols can be pushed, because a large constant increase in the lifetime of a quantum architecture could mean the difference between a physically realistic architecture that can be fault tolerantly operated versus one that cannot, as discussed in Sec. \ref{sec:hybridcodes}.

The rest of the paper is structured as follows: in Sec. \ref{sec:StabCodes}, we review stabilizer codes and how they can be modeled at finite temperature.  In Sec. \ref{sec:ISING}, we describe how the 1D Ising model can be treated as a stabilizer code and discuss the low temperature dynamics of the model.  In Sec. \ref{sec:Protocol}, we construct our autonomous protocol, built out of local unitary operators, and discuss the scaling behavior of the protocol.  We also demonstrate evidence for the enhancement of the lifetime of the 1D Ising model.  In Sec. \ref{sec:TCdynamics}, we sketch how our protocol generalizes to higher dimensions and to other stabilizer codes, including the toric code.
	
\section{Stabilizer Codes}
\label{sec:StabCodes}

\subsection{Definitions}
\label{sec:StabFinBack}
In this section, we briefly review the theory of stabilizer error correcting codes\cite{Gottesman97}.  Given $n$ qubits, a collection of operators ${S_i}$, and $k$ states $|\psi\rangle_i, i=1,..,k$ which span some subspace of the $n$ qubits, let,

\begin{equation}
S_i |\psi\rangle_i = +1 |\psi\rangle_i \label{eq:stabcond1}
\end{equation}
\begin{equation}
[S_i, S_j] = 0 \label{eq:stabcond2}
\end{equation}

for all $i,j$.  Furthermore, suppose there are $m$ \emph{error} operators $E_j, j=1,..,m$, and that for each of them, there exists some operator $S_j$ such that

\begin{equation}
\{E_i, S_j\} = 0
\end{equation}

Stabilizer codes are those collections of states $|\psi\rangle_i$ and operators $\{S_j\}$ which satisfy the above conditions for error operators belonging to some subset of the Pauli group---tensor products of Pauli operators with the identity.  

For example, given three qubits, let $|\psi\rangle_1 = |\uparrow\uparrow\uparrow\rangle$ and $|\psi\rangle_2 = |\downarrow\downarrow\downarrow\rangle$.  Then the set of operators satisfying \eqref{eq:stabcond1} and \eqref{eq:stabcond2} is $\{\sigma_z\sigma_z I, I \sigma_z \sigma_z\}$.  One can easily determine that the set of error operators corresponding to these two \emph{stabilizer} operators is: $\{ I I I, \sigma_x I I, I \sigma_x I, I I \sigma_x, \sigma_x \sigma_x I, \sigma_x I \sigma_x, I \sigma_x \sigma_x\}$.

More transparently, this 3-qubit stabilizer code encodes two protected states.  If some noise source were to apply any single qubit $\sigma_x$ operator, or any two-qubit $\sigma_x^i \sigma_x^j$ operator, measurement of the set of stabilizer operators would indicate the presence of the error.  Furthermore, the code can actually detect \emph{and} correct single $\sigma_x$ errors.  For example, a meaurement result of $-1,+1$ of the stabilizers $\sigma_z \sigma_z I$ and $I \sigma_z \sigma_z$, respectively, indicates either an error on the first qubit or two errors error on the latter two qubits.  For many noise models, the single error situation is much more likely, thus a single $\sigma_x$ operator applied to the first qubit will more often than not return the qubit back into the protected subspace.

\subsection{Active State Preparation versus Dissipative Hamiltonian Engineering}

Here we will refine our discussion by broadly classifying error correcting approaches into (1) state preparation strategies and (2) Hamiltonian engineering strategies.

The target of both strategies is the same: the generation of an encoded stabilizer state.  In state preparation, a stabilizer encoded state is prepared by the application of a sequence of unitaries.  However, ignoring noise sources, the natural Hamiltonian which describes the system is $H = 0$.  The target of such a strategy is generation of the stabilizer state itself.  Implicitly, some sort of active error measurement and correction needs to be performed once the target state is reached.

In contrast, in Hamiltonian engineering approaches, the encoded state is reached by implementing a Hamiltonian on a set of qubits which has a stabilizer encoded state as its ground state.  The stabilizer state is then preserved by keeping a quantum system at a sufficiently low temperature to suppress errors.

Mixtures of these strategies exist.  For example, one could use a Hamiltonian engineering approach to generate a stabilizer encoded state, and then immediately turn off the Hamiltonian once the desired state was reached, preserving the state at further times with active error correction.  Alternatively, one could use Hamiltonian engineering to prepare the state, and then use a combination of dissipation with an additional protocol to detect or correct errors.  We will focus here on this latter strategy.  Specifically, we will be concerned with systems being dissipatively driven towards the ground state of a Hamiltonian which encodes a stabilizer state, and we will build an autonomous error correction protocol to mitigate the ways in which dissipation alone fails to protect the encoded state.

\subsection{Error Correcting Master Equation}

To dissipatively generate a stabilizer code, one forms the system Hamiltonian as the sum of the stabilizer operators for the code of interest, i.e., $H\ =\ -\sum_i S_i$.  This guarantees that the ground state of that hamiltonian will be the encoded subspace.  Furthermore, this ensures that configurations of the system with errors present are excited states.

To model dissipation in such a code, we employ here a Lindblad master equation.  Without loss of generality, but to simplify analysis, we assume that the bath only operates on the system with purely local errors, and that these local errors correspond to the errors of the stabilizer code of interest.  Given this assumption, the dynamics may be described by the Lindblad equation:

\begin{align}
\dot{\rho }=\sum_{\omega }{2c_{\omega }\rho c^{\dagger }_\omega}-c^{\dagger }_{\omega }c_{\omega }\rho -\rho c^{\dagger }_{\omega }c_{\omega }, \label{eq:Lindblad}
\end{align}

where $\rho$ is the system density matrix for some candidate system and $\{c_\omega\}\ =\ \{ \sqrt{\gamma_\omega} L_\omega \}$ are Lindblad operators arising from interactions with a bath, where the $L_\omega$ act on the system with characteristic rates $\gamma_\omega$.  Error processes can then be represented by products of the Lindblad operators: $\{c_\omega^1 c_\omega^2 \cdots c_\omega^n \}$.

A necessary condition for error correction to occur to $n^{\rm{th}}$ order in the error processes is to apply the inverses of the error processes sufficiently rapidly.  If we restrict ourselves to stabilizer codes on lattices, then the recipe for error correction is straightforward: measure the stabilizers of the code and apply correction operations conditioned on the results of the stabilizer measurements.

While it is in principle possible to measure all of the stabilizers of a given system simultaneously because they all commute, it will be convenient to decompose a given correction protocol into groups of terms involving operators only acting within a characteristic length scale $\lambda$.  This is useful because it provides a natural scale for treating stabilizer codes with fixed resources, and it allows the interpretation of different protocols as the implementation of a certain kind of effective long-range interaction.

\subsection{Error Correction Thresholds and Scaling}

Much of the power of stabilizer codes arises from the existence of error thresholds.  Specifically, as the stabilizer code is made sufficiently large, the probability of remaining in an encoded subspace goes to 1 as long as measurement/correction cycles occur faster than the threshold rate.  This gives rise to a competition between the resources necessary to perform error correction/detection for stabilizer codes involving many qubits, versus the scaling of the error rate of the code with system size.  

\begin{figure}
\begin{center}
\scalebox{1}{\includegraphics[width=1.0\columnwidth]{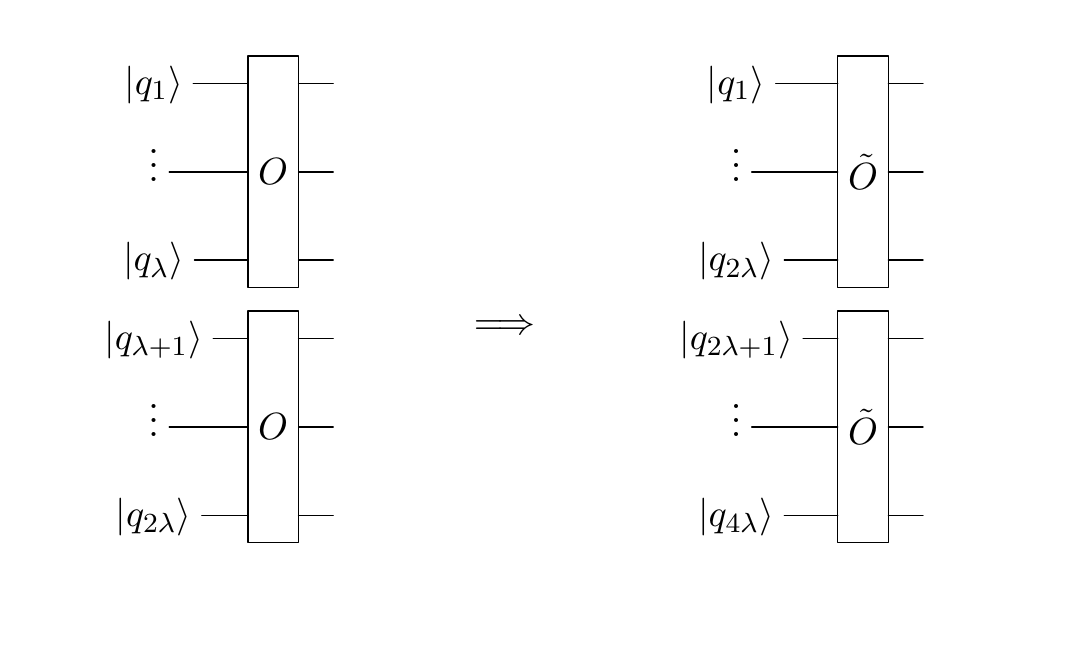}}
\end{center}
\caption{A linear stabilizer code with detection/correction operators $O$.  When the code is made larger, correction of errors in the code will require measurement and correction of errors occurring over larger lengthscales.}
\label{fig:renormcorrection}
\end{figure}

For concreteness, consider a linear stabilizer code, with correction/detection steps idealized by operators $O$ acting over a length scale $\lambda$ as in \figref{fig:renormcorrection}.  In reality, these operators $O$ can often themselves be decomposed into purely local operators, but detecting and correcting errors occurring over a length scale $\lambda$ requires measuring and applying many such local operators over that length scale sufficiently quickly.  For sufficiently large $\lambda$ and fast application of gates, the distinction in terms of resource requirements between a nonlocal operator $O$ acting in a region $\lambda$ versus a sequence of local operators acting within a region $\lambda$ becomes a matter of philosophy.

Thus, ensuring the code remains in a protected subspace as system size is made large can, in practice, require applying sequences of operators over successively larger length scales.  Generically, having a larger stabilizer code provides better protection to errors, but this must be weighed against the growth rate of the number of gates necessary to implement the corrective operators $O$ as system size and $\lambda$ are varied.  Note that if an error correction threshold exists for the code, the number of gates could, in principle, stop growing.  But, in the absence of a threshold, if gates can only be applied with rate $\chi$, the maximum system lifetime is set by the scaling properties of $O$.  We explicitly construct this operator for a special case in Sec. \ref{sec:ECDiss}.

\section{1D Ising Model}
\label{sec:ISING}

\subsection{1D Ising Model as a Stabilizer Code}

The choice of 3-qubit stabilizer code introduced in \ref{sec:StabFinBack} was deliberate, because it can naturally be extended and interpreted as the ground state of a 1-dimensional Ising model.

\begin{equation}
H_{1D Ising} = -\Delta \sum_{i=1}^{L} \sigma_{z}^{i} \sigma_{z}^{i+1} \label{eq:IsingHam}
\end{equation}

The ground state subspace of this model is two-fold degenerate and is comprised of the states $|\uparrow\cdots\uparrow\rangle$ and $|\downarrow\cdots\downarrow\rangle$.  These ground states are exactly the $L$-qubit analogues of the 3-qubit code treated previously.  These states are stabilized by the set of all adjacent pairwise $\sigma_z$ operators $\{ I I  \cdots I \sigma^i_z \sigma^{i+1}_z I \cdots I \}$, where $i$ runs from $1$ to $L$.  These are precisely the operators appearing in the Hamiltonian of the 1D Ising model.  For the remainder of our analysis, we assume without loss of generality that $\Delta=1$.

Furthermore, $\sigma_x$ errors are equivalent to excited states.  In the simplest case, errors can be corrected by resorting to a simple majority rule---if most spins point in a particular direction, the correction protocol returns the state to the encoded ground state corresponding to that direction.

\subsection{1D Ising Model at Finite Temperature}

By coupling the Ising model to an external reservoir, one might hope to dissipatively drive the 1D Ising model into one of these encoded states.  However, the 1D Ising model has no finite temperature ordered phase, so at all finite temperatures, the system evolves towards the thermal state.  Furthermore, this timescale over which the system relaxes to a thermal state is known to be independent of the size of the chain, given modest bath assumptions\cite{Glauber1963d}.  Thus, dissipation by itself cannot protect the 1D Ising model, and an additional protocol needs to be implemented in order to correct thermal errors. While dissipation cannot protect the 1D Ising Model at finite temperature, it is instructive to understand the details of how thermal fluctuations lead to instability in this simple case, because very similar processes are responsible for the instability of many other stabilizer codes at finite temperature. In previous work, we examined the dynamics of this model, as well as of the toric code, at finite temperature\cite{Freeman2014}.  In particular, we identified a low temperature regime where the dynamics are well described by a simple random walk model.  We briefly summarize the analysis below.

When studying the error dynamics, it is convenient to consider the dual lattice of the Ising model: we imagine a new 1D lattice with sites interleaved between the sites of \eqref{eq:IsingHam} and associate auxiliary spin values $b_i$ with them.  The auxiliary site's spin values are uniquely determined by the products $b_i = \sigma_{z}^{i} \sigma_{z}^{i+1}$, where site $b_i$ defined by this equation sits between site $i$ and $i+1$.  We can identify these extra variables with \emph{domain walls}.  If adjacent spin variables disagree, then the auxiliary site sitting between them will have $b_i=-1$.  If all but a contiguous block of spins disagree, then all auxiliary sites will have $b_i=1$ except for those two sites which sit at the two boundaries of the contiguous blocks of spins.  Describing the dynamics of these domain walls is equivalent to describing the spin dynamics, because if one knows all the auxiliary variables plus any single spin value, $\sigma_z^i$, one can reconstruct all of the remaining spin variables $\sigma_z^j$.

For simplicity, we assume a bath that operates on the system only by creating, destroying, or translating domain walls.  Then, for sufficiently low temperatures, occasionally the bath will cause an adjacent domain wall pair to appear in the system.  Bath fluctuations will cause this pair of domain walls to fluctuate across the system, effectively causing the domain walls to undergo a 1D random walk.  When domain walls are adjacent, it is energetically favorable for them to be dissipated.  If domain walls fuse \emph{before} traversing the length of the system, the encoded state will be preserved.  But if domain walls undergo a random walk such that one winds entirely around the system, this effectively performs an uncorrectable error on the encoded qubit because the system will have transitioned from one encoded ground state to the other encoded ground state\cite{Freeman2014}.

\subsection{Microscopic Master Equation}

When the bath operates on the system with purely local errors which only create, destroy, and translate domain walls, the Lindblad operators for are of the form:

\begin{equation}
\left \{ c_\omega \right \} = \left \{ \sqrt{\gamma_0} T_{b}, \sqrt{\gamma_+} D^\dagger_{b}, \sqrt{\gamma_-} D_{b} \right\} 
\end{equation}

When resolved in the Pauli basis, these operators take a simple form:

\begin{align}
D^\dagger_{b}&= \frac{1}{4} \left(I  \sigma_x  I + \sigma_z  \sigma_x  \sigma_z + i \left(I  \sigma_y  \sigma_z + \sigma_z  \sigma_y  I \right) \right), \notag \\ 
D_{b}&= \frac{1}{4} \left(I  \sigma_x  I + \sigma_z  \sigma_x  \sigma_z - i \left(I  \sigma_y  \sigma_z + \sigma_z  \sigma_y  I\right)\right), \notag \\
T_{b}&=\frac{1}{2} \left(I  \sigma_x  I - \sigma_z  \sigma_x  \sigma_z \right) \label{eq:LindbladDef}
\end{align}

A short calculation verifies $\sum_i c_i c^\dagger_i = I$.  Physically, these operators represent the creation of a domain wall pair at dual lattice sites $b$ and $b+1$ ($D^\dagger_{b}$), annihilation of a pair of domain walls at dual lattice sites $b$ and $b+1$ ($D_{b}$), and the translation of a domain wall from $b$ to $b+1$ or $b+1$ to $b$ ($T_{b}$).  Additionally, these operators only connect diagonal elements of the density matrix to other diagonal elements.  This reduces the time evolution of the diagonal matrix elements to a classical master equation:

\begin{align}
\frac{d P_n}{d t} = \gamma_0 \sum_{n_0}  \left( P_{n_0}-P_n \right)&+ \sum_{n_+}  \left( \gamma_- P_{n_+}- \gamma_+ P_n \right) \notag \\
&+ \sum_{n_-} \left( \gamma_+ P_{n_-} - \gamma_- P_n \right) \label{eq:ME}
\end{align}

The rates which which these operators are applied, i.e. $\gamma_0, \gamma_+,$ and $\gamma_-$, are set by the specific choice of bath model.  For simplicity, we consider here a Markovian bath.  The rates of such a bath are determined by:

\begin{align}
\gamma \left( \omega \right)=\xi \left  \vert \frac{\omega^n }{1-e^{-\beta \omega }} \right \vert 
\end{align}

The relevant rates for our study are $\gamma_-$, $\gamma_+$, and $\gamma_0$, corresponding to domain wall pair annihilation, pair creation, and translation, respectively.  Different $n$ correspond to different types of baths--for $n=1$ the bath is Ohmic, and for $n\geq2$, the bath is Superohmic.  For our purposes, it will be more convenient to treat $\gamma_0$ as a tunable parameter to study the scaling behavior of our protocol.  Qualitatively, $\gamma_0$ scales linearly with $T$ for Ohmic baths and equals zero for Superohmic baths.  For simplicity, we work in units where $\xi=1$.

For more details of the master equation approach used to study this model, see Ref.~\onlinecite{Freeman2014}.

\subsection{Error Correcting Operator}
\label{sec:ECDiss}

\begin{figure}
\begin{center}
\scalebox{1}{\includegraphics[width=1.0\columnwidth]{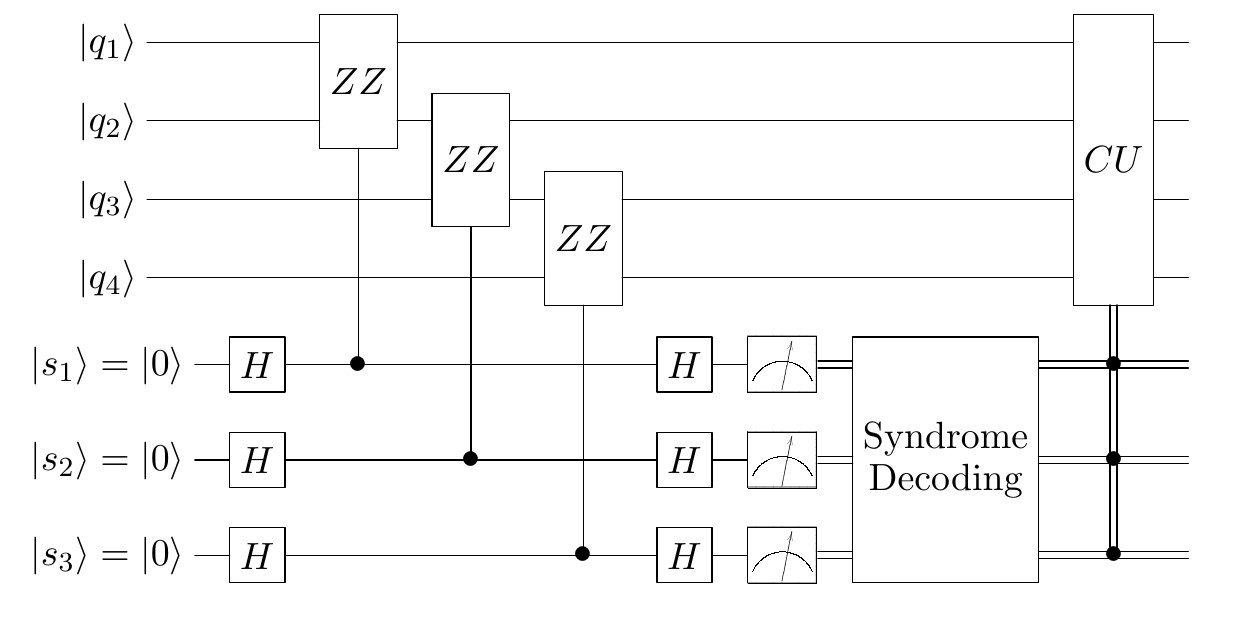}}
\end{center}
\caption{Circuit for performing error suppression for a subregion of the 1D Ising model.  Qubits $q_1$ through $q_4$ are lattice sites on an Ising chain.  $s_1$ through $s_3$ are ancillas used to read out the syndrome measurements of $ZZ$ on the nearest-neighbor Ising lattice sites.  Based on the results of the syndrome measurements, the conditional unitary operator $CU$ corrects the errors present.  A table which defines $CU$ is provided in Appendix \ref{sec:syndecode}.  This entire circuit represents a possible realization of one such operator $O$ from \figref{fig:renormcorrection}.}
\label{fig:lambda3ECC}
\end{figure}

In the absence of resource constraints, it is straightforward to construct the operators which correct errors in the 1D Ising model.  According to the schematic shown in \figref{fig:renormcorrection}, the $\lambda=2$ analogue of $O$ is simply the domain-wall annihilation operator, $D_b$ from \eqref{eq:LindbladDef}.  In more generality, for larger $\lambda$ the corresponding $O$ is the operator which, given an even number of domain walls, annihilates all domain wall pairs in the region being operated upon.  For example, the circuit for the $\lambda=3$ version of this operator is depicted in \figref{fig:lambda3ECC}.  Note that for an odd number of domain walls, there is not an unambiguous choice for how to annihilate domain walls because a free, unpaired domain wall is always left over.

\section{The Protocol}
\label{sec:Protocol}

\subsection{Protocol Considerations for the 1D Ising Model}

The most straightforward error correcting protocol is simply to measure the system's stabilizers often enough that one can unambiguously locate pairs of domain walls and then perform correction operations, as indicated in \figref{fig:lambda3ECC}.  This can be represented by a sequence of measurement operators, the stabilizer for the 1D Ising model, $S_i$, interleaved by conditional application of corrective unitaries: $\rm{DSWAP}$ and $\rm{DWALL}$.  These operators operators have the following representation in the Pauli basis:

\begin{align}
\rm{DWALL} &= \frac{1}{2} \left(III + I\sigma_x I - \sigma_z I \sigma_z + \sigma_z \sigma_x \sigma_z \right) \\
\rm{DSWAP} &= \frac{1}{2} \left(III + I\sigma_x I + \sigma_z I \sigma_z - \sigma_z \sigma_x \sigma_z \right)
\end{align}

where there is a pair of these operators for each triple of lattice sites.  In \figref{fig:lambda3ECC}, the details of the 4-qubit operator $O$ are abstracted away (see Appendix \ref{sec:syndecode}), but it can be decomposed into applications of $\rm{DSWAP}$s and $\rm{DWALL}$s, conditioned on syndrome measurements.  A simple calculation shows $[H_{Ising},\rm{DSWAP}]=0$ and $[H_{Ising},\rm{DWALL}]=+1$.

Intuitively, $\rm{DWALL}$ destroys (creates) a domain wall pair at the dual lattice site in-between the three qubits being operated on if and only if a domain wall pair is present (or, all of the spins are aligned), respectively.  \rm{DSWAP} translates a domain wall, either left or right if and only if a single domain wall exists between the 3 spins being operated on.

Because we seek a protocol \emph{without} measurements, the natural operators for such a procedure are \rm{DWALL} and \rm{DSWAP}.  \rm{DWALL} is inconvenient, both because the bath already acts to dissipate excitations and because it can lead to the generation of extra, uncontrolled domain walls more easily than the \rm{DSWAP} operator.  Consequently, we only use \rm{DSWAP}s in our protocols.

If we restrict our attention to the low temperature regime, then the lifetime of the Ising chain is governed by the dynamics of single pairs of defects.  For error correcting purposes, it is convenient to classify the common geometries of pairs of domain walls.  First, \emph{correctable errors} are those errors for which the pair of domain walls is not yet separated by $L/2$ or more.  \emph{Non-correctable} errors are those domain wall configurations in the complement of this set.  In the language of error correction, the distance for this code is $\lfloor L/2\rfloor$--more transparently, correctable errors are those errors which will be correctly matched by a perfect decoder.  Furthermore, we need to distinguish between \emph{trivial} and \emph{nontrivial} defect pairs.  A domain wall pair is trivial if two domain walls sit on neighboring dual lattice sites.  Again, assuming we operate in the low temperature regime, these trivial defect pairs annihilate with rate $\gamma_-$---that is, much faster than other time scales of the problem.  Nontrivial pairs are those pairs which are not on neighboring dual lattice sites.  

Designing a successful protocol for the Ising model amounts to designing a sequence of \rm{DSWAP}s that efficiently dissipates nontrivial, correctable defect pairs.  If we let $\chi$ be the rate at which DSWAPs can be applied, then we expect an enhanced lifetime given the following rate assumptions:

\begin{equation}
\gamma_- >>  \chi / O(\rm{poly}(L))  \sim \gamma_0 > \gamma_+.
\end{equation}

To wit, DSWAPs are applied at a rate much slower than the inherent annihilation rate of the system--this is so DSWAPs do not turn trivial defect pairs into nontrivial defect pairs.  Furthermore, $\chi$ is chosen to be close to the inherent translation rate so that correctable, nontrivial defect pairs can be brought adjacent to one another and then be dissipated by the bath before they have time to translate out of the correctable range of the protocol.  The $O(poly(L))$ factor multiplying $\chi$ accounts for the fact that different protocol require some polynomial in $L$ number of swaps to sweep across the entire lattice.  For a proof of the polynomial scaling in $L$, see Appendix \ref{sec:MATCHSEQ}.

In the absence of a corrective protocol, this intrinsic hopping rate of the Ising model gives rise to a simple, background error rate\cite{Glauber1963d, Freeman2014},

\begin{equation}
\Gamma_{0}=\frac{\gamma_0}{1+e^{\Delta/T}} \label{eq:isingglauber}
\end{equation}

\subsection{Protocol Construction}

In this section we construct an autonomous error correction protocol for the 1D Ising model with a variable length-scale $\lambda$.  The design of the protocol reduces to attempting to perform a sequence of $\rm{DSWAP}$s that will necessarily cause any arbitrarily placed pair of domain walls within a region of length 2$\lambda$ to become neighbors.  We refer the reader to Appendix \ref{sec:MATCHSEQ} for a more complete discussion of this strategy.

There are a variety of ways to construct protocols which achieve this in a number of $\rm{DSWAP}$s that scales polynomially in the length of the system.  Here we focus on protocols which we call $\lambda$-mixing.  By definition, these are protocols which, in the absence of errors, never translate domain walls a distance $\lambda$ or greater. For an Ising model of length $L$, $\lambda$ runs from $1$ to $\left \lfloor{L/2}\right \rfloor$.  In the language of error correction, the protocol can be designed to correct errors of distance $1$ to distance $\left \lfloor{L/2}\right \rfloor$.

\begin{figure}
\begin{center}
\scalebox{1}{\includegraphics[width=1.0\columnwidth]{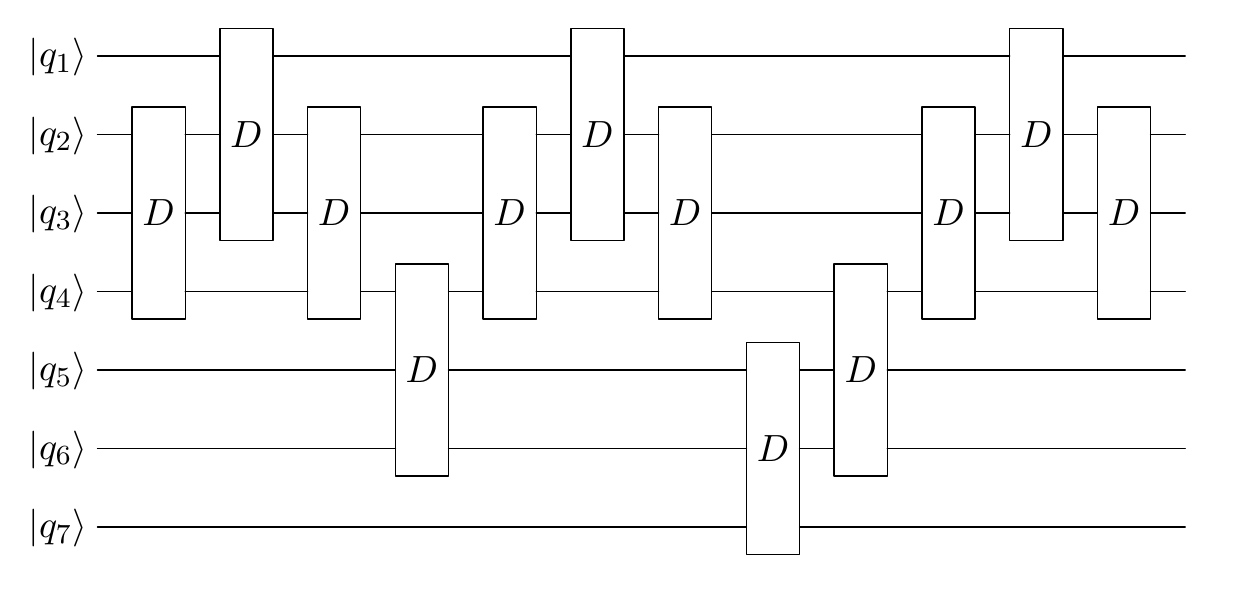}}
\end{center}
\caption{Sequence of $\rm{DSWAP}$s, denoted $C$, for a $\lambda=3$ $\lambda$-mixing protocol.  If a pair of domain walls exist anywhere between sites $q_1$ through $q_7$, then they will necessarily be brought adjacent to each other by this sequence of swaps.  Gates are applied sequentially with waiting time $1/\chi$ between each gate.}
\label{fig:lambda3circuit}
\end{figure}

\begin{figure}
\begin{center}
\scalebox{1}{\includegraphics[width=1.0\columnwidth]{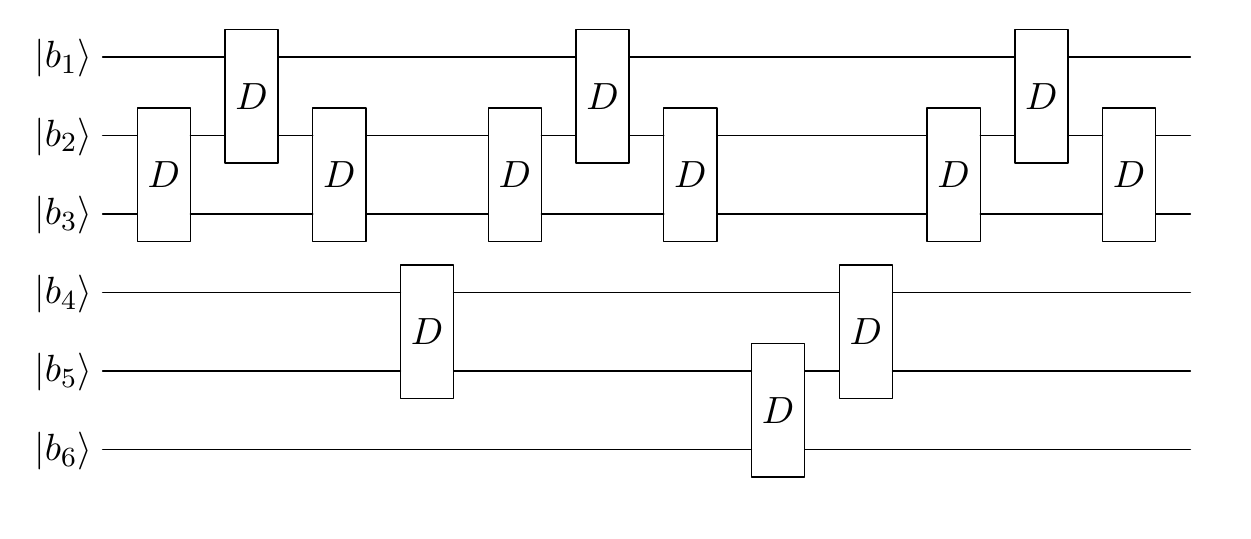}}
\end{center}
\caption{The same sequence from \figref{fig:lambda3circuit} but shown acting on domain-wall variables.  Here, it is clear that the sequence of $\rm{DSWAP}$s is designed not to mix domain walls between the two regions of size $\lambda=3$.  Site $b_1$ sits between $q_1$ and $q_2$, $b_2$ between $q_2$ and $q_3$, etc.}
\label{fig:lambda3circuitdomains}
\end{figure}

\begin{figure}
\begin{center}
\scalebox{1}{\includegraphics[width=1.0\columnwidth]{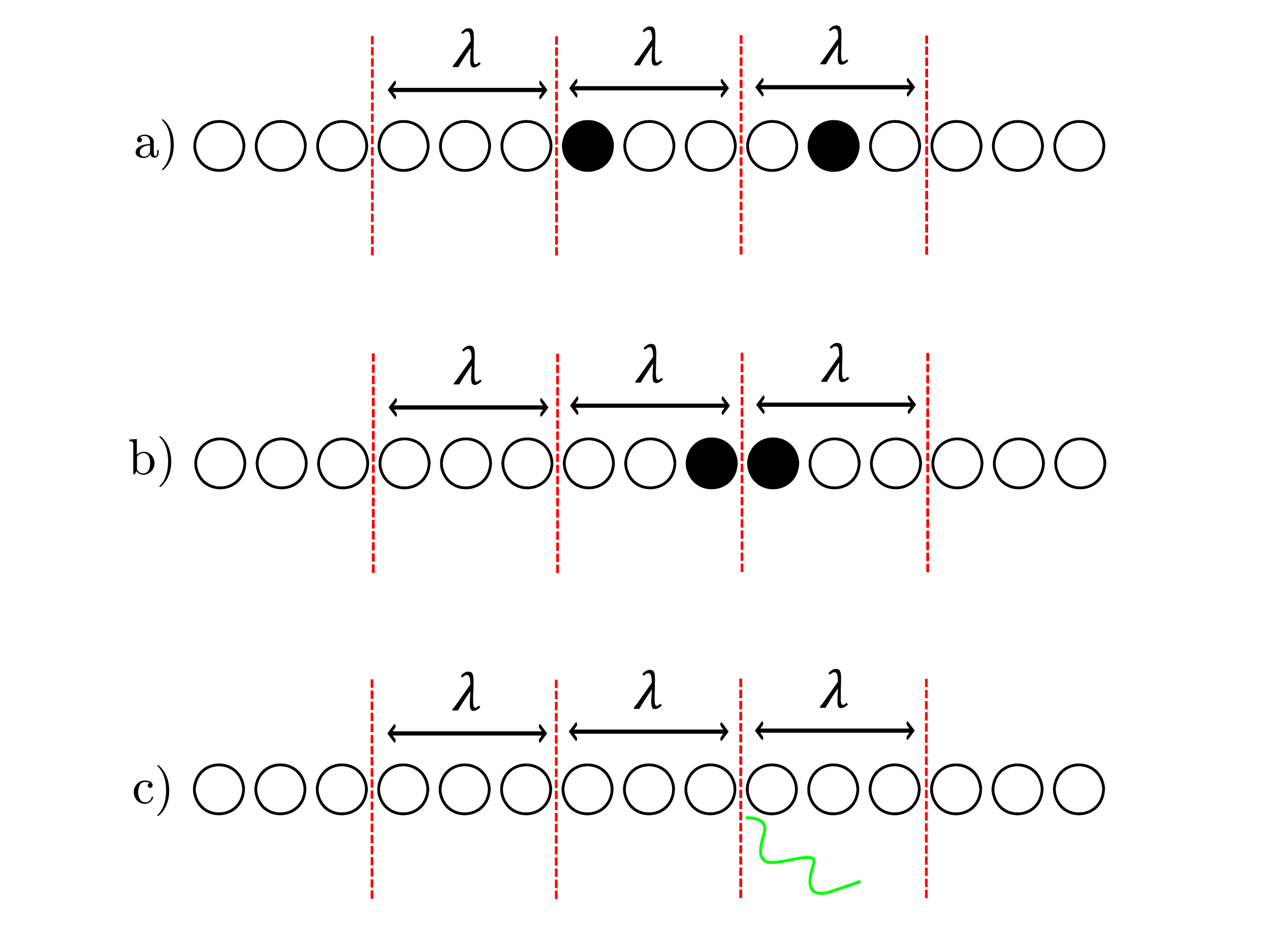}}
\end{center}
\caption{One possible snapshot of the error correction process.  $a$) depicts a system with two domain walls present, each sitting in adjacent $\lambda$-domains. $b$) depicts the state of the system after the protocol has been applied--domain walls have been shuttled to the shared boundary.  In $c$), the bath dissipates the domain walls, and the system returns to the ground state.}
\label{fig:defectcorrection}
\end{figure}

First, the dual lattice is subdivided into non-intersecting subregions of length $\lambda$.  Then, two adjacent regions are chosen, and a $\lambda$-mixing protocol is applied over that subregion of total length $2\lambda$.  DSWAPs are chosen to move defects towards the shared boundary of the two regions, but not to mix defects between the boundaries.  The non-intersection of the two regions is crucial: if the protocol did not have this feature, it would actually \emph{increase} the error rate, effectively increasing the inherent translation rate, and thus diffusion rate of defects in the system.   \figref{fig:lambda3circuit} depicts a circuit for this protocol for $\lambda=3$ and \figref{fig:lambda3circuitdomains} depicts the same circuit acting on the domain wall variables.  \figref{fig:defectcorrection} illustrates a snapshot of this entire procedure for a representative error process involving two domain walls sitting in neighboring $\lambda$-domains.

This circuit should be reminiscent of the cartoon sketched in \figref{fig:renormcorrection}.  For our purposes, the operator $O$ from \figref{fig:renormcorrection} is the full sequence of $\rm{DSWAP}$s in \figref{fig:lambda3circuit} dressed by the probabilistic action of creation/annihilation/translation operators by the bath on the system.

We provide code for this algorithm in the Appendix \ref{sec:Algorithm}, including how the $\lambda$-mixing subprotocols are constructed.

\subsection{Error Modes and Scaling}

In this section, we examine how uncontrollable thermal errors lead to loss of the qubit in the presence of the protocol.

In the presence of a corrective protocol, and assuming the correction rate $\chi$ is close to the translation rate of the system but still much less than the annihilation rate, the lowest order error process is,

\begin{equation}
\Gamma_{\rm{Cyc}} = L \gamma_+ \frac {\gamma_0}{\lambda \chi} \frac{\gamma_0}{\gamma_-} \frac{1}{L-2-2\lambda} \frac{f(\lambda)}{L} \label{eq:lowchidynamics}
\end{equation}

This rate is the product of (i) the baseline production rate of defect pairs, $L \gamma_+$, (ii) the probability of a defect pair not immediately annihilating, $\frac{\gamma_0}{\gamma_-}$, (iii) the probability of a defect exiting a corrective region, $\frac{\gamma_0}{\lambda \chi}$, (iv) the probability of a nontrivial random walk across the chain $\frac{1}{L-2-2\lambda}$, divided by a factor proportional to the number of correcting regions on the lattice.  Thus, for fully parallel application $f(\lambda) \propto \lambda$.  Without the protocol, the probability that a pair of domain walls undergoes a random walk that winds around the entire system scales like $\frac{1}{L-2}$, but when the protocol is implemented, the effective lattice size is slightly reduced: the particle need only come within approximately a distance $2\lambda$ of its partner for the protocol to fuse them.

This effective rate is valid as long as $\chi$ is fast enough to compete with $\gamma_0$, but not so fast as to compete with pair annihilation, $\gamma_-$, and other higher order processes in $\frac{\gamma_0}{\gamma_-}$ and $\frac{\gamma_0}{\chi}$.  It might be tempting to examine the form of \eqref{eq:lowchidynamics} and expect that errors vanish in the limit of $\gamma_0 \rightarrow 0$, but a new effective translation rate appears once $\gamma_0 << \gamma_+$.  In this regime, two pairs of domain walls can appear next to one another, and a consecutive annihilation event produces a lone of pair of domain walls separated by two dual-lattice sites.  In this way, an effective translation rate is set by the rate at which these doubled-pair creation events occur.  We do not consider this limit further, but it is the natural error process for superohmic baths at low temperature.

To model the breakdown of \eqref{eq:lowchidynamics} as $\chi$ is varied, we can approximate the lifetime, $\frac{1}{\Gamma_{\rm{Cyc}}}$, as being effectively reduced by some factor proportional to $\chi$: 

\begin{equation}
\frac{1}{\Gamma_{\rm{Cyc}}} \rightarrow \frac{1}{\Gamma_{\rm{Cyc}}} (1 - \chi g(\lambda, L, \gamma_0,\gamma_-,\gamma_+) + O((\frac{\chi}{\gamma_-})^2+(\frac{\chi}{\gamma_0})^2), \label{eq:rateseqn2ndorder}
\end{equation}

with $g(\lambda, L)$ a protocol-dependent scaling function.  Heuristically, for fixed $\lambda$, one expects that $g$ should scale linearly with the number of parallel domains of size $\lambda$ because, for twice as many domains, twice as many pairs will be pulled apart by the protocol that would have otherwise fused.  At the same time, for a fixed number of domains, i.e. fixed $\frac{L}{\lambda}$, any given pair of lattice sites is only ever operated on by a DSWAP for a fraction of the corrective cycle.  So, for fixed $\chi$ and fixed $\frac{L}{\lambda}$, as $\lambda$ is increased, domain walls may spend a longer amount of time sitting on a boundary before being caught by the protocol.  For the protocol used in this paper, this is cubic in $\lambda$.  Thus,

\begin{equation}
g(\lambda, L, \gamma_0,\gamma_-,\gamma_+) \propto g(\gamma_0,\gamma_-,\gamma_+) \lambda^3 \frac{L}{\lambda} = g(\gamma_0,\gamma_-,\gamma_+) \lambda^2 L
\end{equation}

This scaling behavior suggests a critical cycling rate, $\chi_c$, at which the lifetime is maximally improved by the protocol.  Differentiating \eqref{eq:rateseqn2ndorder} with respect to $\lambda$ yields the critical rate, up to the rate function $g$,

\begin{equation}
\chi_c = \frac{1}{2 \lambda^2 L g(\gamma_0,\gamma_-,\gamma_+)},  \label{eq:chicrit}
\end{equation}

where any residual prefactors and terms involving $\gamma_0$, $\gamma_+$, and $\gamma_-$ have been absorbed into $g$.

\subsection{Memory Enhancement and Scaling}

We now present numerical results demonstrating the enhanced lifetime of the Ising Model when subjected to $\lambda$-mixing protocols in serial and in parallel.  For serial application, only a single corrective operation was applied every $1/\chi$ units of time.  For parallel application, $L/(2\lambda)$ simultaneous corrective operations were applied every $1/\chi$, where each operation acted on a nonintersecting region of length $(2\lambda)$.

For the following analysis, we define the \emph{lifetime} as the average time it takes a 1D Ising model initialized to the spin up state to transition to the spin down state.  In the absence of the protocol, that is, in the low-$\chi$ limit, this lifetime asymptotes to approximately the lifetime given by \eqref{eq:isingglauber}.

For the details of the Monte Carlo algorithm, see Ref.~\onlinecite{Freeman2014}.  The only nontrivial choice required at the level of simulation is how to treat the competition between the application $\rm{DSWAPs}$ and bath operators.  For simplicity, we assume if a bath operator takes longer than $1/\chi$ to occur, that the DSWAP occurs unhindered.  Likewise, if a bath operator takes less than $1/\chi$ to occur, the transformation associated with that bath operator occurs unhindered, be that a pair creation, pair annihilation, or single translation.  More complicated choices could be made, like choosing a probabilistic failure rate of a $\rm{DSWAP}$ as a function of the ratio of the competing timescales, but we do not expect the result of a such a treatment to greatly affect our analysis.

\begin{figure}
\begin{center}
\scalebox{1}{\includegraphics[width=1.0\columnwidth]{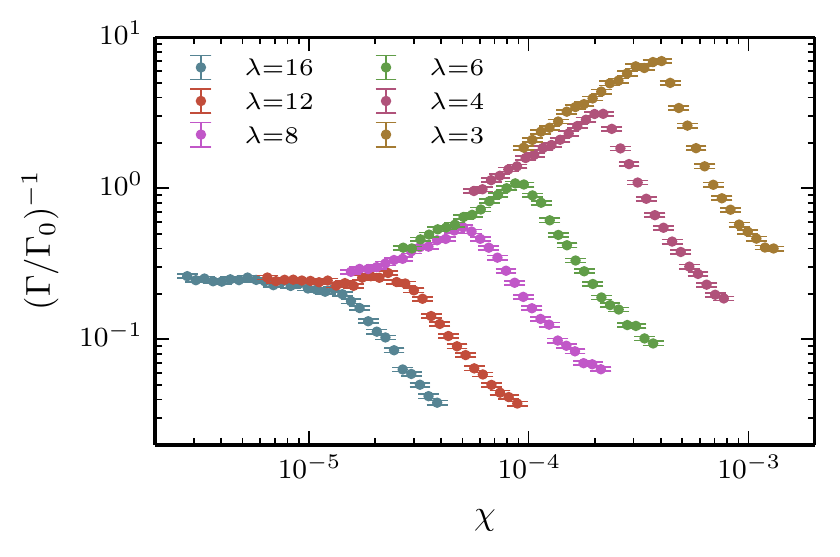}}
\end{center}
\caption{Lifetime of the Ising model, expressed in terms of inverse units of $\Gamma_{Ising}$  for a range of $\chi$ and for different values of $\lambda$ for $L=96$, $T=.07$, $\gamma(0) = .0007$.  Protocols were implemented in parallel on $48/\lambda$ blocks (see text).  In the absence of the protocol, the lifetime of the Ising model for these parameters corresponds to approximately $\Gamma_{0}^{-1}$, i.e. \eqref{eq:isingglauber}.  This is the value which all three protocols converge towards in the limit of $\chi << \gamma_0$.  Note the decrease in lifetime for $\chi \approx \gamma_- = 1$.}
\label{fig:parallelmemscaling}
\end{figure}

\figref{fig:parallelmemscaling} depicts the scaling of the 1D Ising model's lifetime with $\lambda$ at fixed $L$, where a smaller $\lambda$ results in more domains being operated on in parallel.  Specifically, for parallel simulations, the protocol was performed simultaneously on $L/(2\lambda)$ domains.  These domains were chosen such that $\rm{DSWAP}$s were only being applied on non-overlapping regions of characteristic size $\lambda$.  Here, increasing parallelization manifestly increases the lifetime of the model.  For small $\chi$, the protocol does nothing, and the memory converges to the value of the memory in the absence of any corrective protocol, i.e. \eqref{eq:isingglauber}.  For $\chi$ approaching $\gamma_-$, the protocol begins to compete with the process of pair annihilation, and begins turning trivial defect pairs into nontrivial pairs.  This actually reduces the lifetime below that of the protocol-free value.  In the intermediate regime, the optimal lifetime grows linearly with the number of parallel blocks employed in the algorithm.  For this particular protocol, the number of parallel blocks was $48 / \lambda$.

\begin{figure}
\begin{center}
\scalebox{1}{\includegraphics[width=1.0\columnwidth]{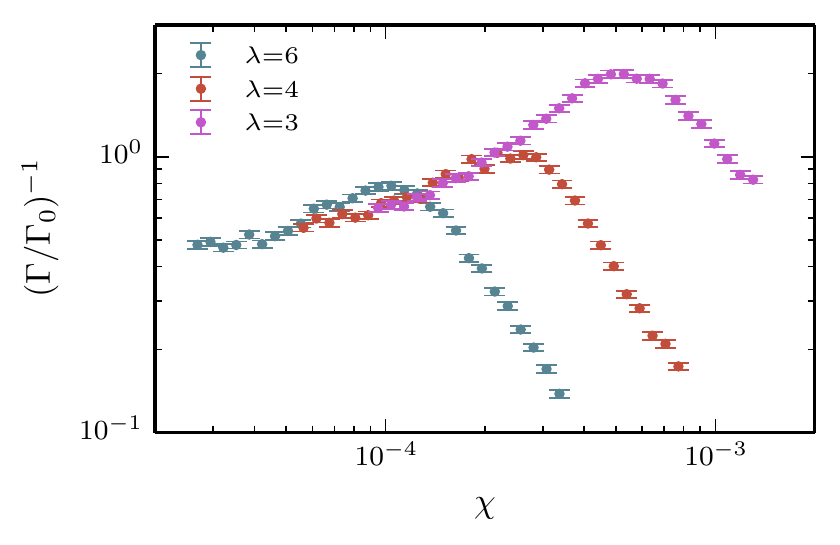}}
\end{center}
\caption{Lifetime of the Ising model for a range of $\chi$ and for different values of $\lambda$ for $L=96$, $T=.07$, $\gamma(0) = .0007$.  Protocols were implemented serially (see text).  The scaling of lifetime with $\chi$ is characteristically similar to the parallel case; however, the maximal lifetime is correspondingly smaller for the serial implementation.  Note that smaller $\lambda$ still yields a larger enhanced lifetime.}
\label{fig:serialmemscaling}
\end{figure}

\figref{fig:serialmemscaling} depicts the scaling of lifetime with $\lambda$, as in \figref{fig:parallelmemscaling}, but for a serial application of the protocol.  For serial application, only a single $\rm{DSWAP}$ operator ever operates on the system over a timescale $\chi^{-1}$.  Decreasing $\lambda$ also manifestly increases the maximum enhanced lifetime of the protocol.  Thus, for fixed-resource architectures, smaller $\lambda$ necessarily outperforms larger $\lambda$ implementations. 

\begin{figure}
\begin{center}
\scalebox{1}{\includegraphics[width=1.0\columnwidth]{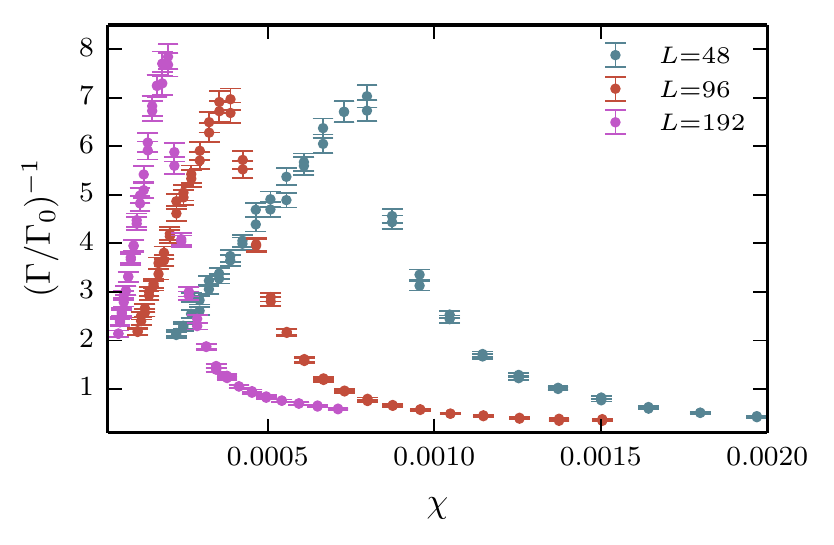}}
\end{center}
\caption{Lifetime of the Ising model for a range of $\chi$ and for different values of $L$ for $\lambda=3$, $T=.07$, $\gamma(0) = .0007$.  Protocols were implemented in parallel on $L/2\lambda$ blocks (see text).  Note the linear scaling in $\chi$ for small values, as well as the shift in the maximum of the lifetime as a function of $L$.}
\label{fig:parallelsizescaling}
\end{figure}

\begin{figure}
\begin{center}
\scalebox{1}{\includegraphics[width=1.0\columnwidth]{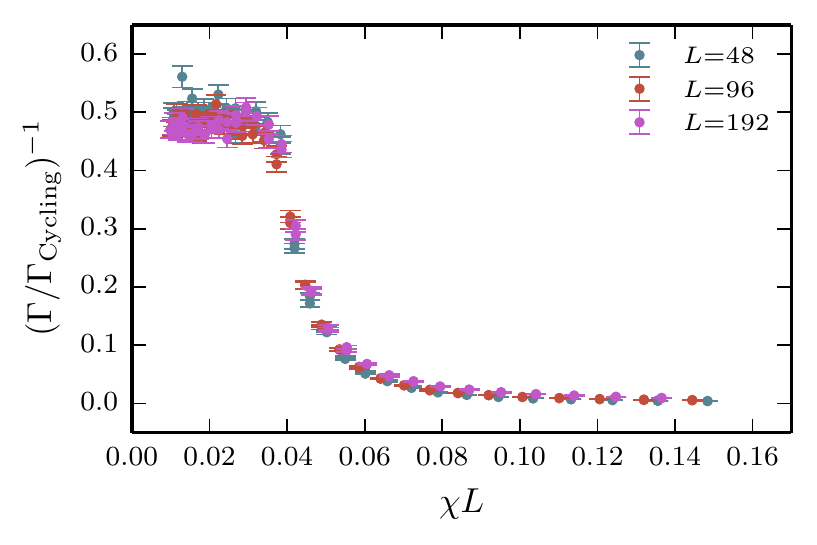}}
\end{center}
\caption{This figure contains the same data as \figref{fig:parallelsizescaling}, but with the $\chi$ axis rescaled to $\chi L$, and the $1/\Gamma_{\rm{Cyc}}$ axis rescaled by \eqref{eq:lowchidynamics}.  Hence, the linear scaling in $\chi$, and the slight residual system size dependence have been removed.  Note the steep, sudden dropoff in lifetime after $\chi L \sim .03$.}
\label{fig:pssdatacollapse}
\end{figure}

\figref{fig:parallelsizescaling} and \figref{fig:pssdatacollapse} depict the scaling of the lifetime with $L$ at fixed $\lambda$ for parallel application.  Remarkably, the $L$ dependence of the models can be completely removed by rescaling the data by \eqref{eq:lowchidynamics}, and rescaling $\chi$ to $\chi L$ as depicted in \figref{fig:pssdatacollapse}.  This rescaling reveals the turnaround in the scaling of the lifetime for $\chi L = .038 \pm .002$, whereafter it transitions from linear scaling in $\chi$ to a power law decay.
 
\begin{figure}
\begin{center}
\scalebox{1}{\includegraphics[width=1.0\columnwidth]{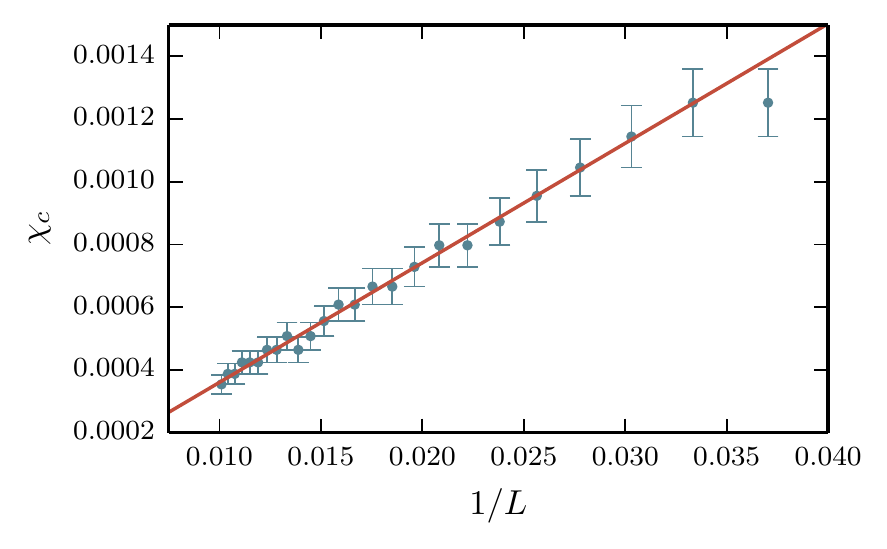}}
\end{center}
\caption{The critical cycling rate, $\chi_c$ as a function of $\frac{1}{L}$ for $\lambda=3$, $T=.07$, $\gamma(0) = .0007$.  Protocols were implemented in parallel on $L/2\lambda$ blocks.  This scaling is consistent with the error model ansatz in \eqref{eq:chicrit}.  Fit to $1/L$ in red.  Errors are dominated by systematic effects, not sampling error.}
\label{fig:chicritical}
\end{figure}

\section{Higher Dimensions and Generalization}
\label{sec:TCdynamics}

\subsection{The Toric Code}

The argument and construction from the previous section immediately generalizes to any higher dimensional stabilizer codes with stringlike error operators. The immediate analogue is Kitaev's Toric Code, whose Hamiltonian is defined as a sum over vertex and plaquette operators acting on the edges of a square lattice,

\begin{align}
\Htc &= -J_e \sum_v A_v -J_m \sum_p B_p ,\label{eq:HTC}\\
A_v &\equiv \prod_{j \in v} \sigma_j^z,\quad B_p \equiv \prod_{j \in p} \sigma_j^x,\label{eq:AvBp}
\end{align}

The low temperature dynamics of the toric code are governed by the proliferation of localized excitations that are created by string-like error operators, with dynamics similar to to those of the 1D Ising model.  However, toric code dynamics differ in two ways: first, there are now two types of defects in the toric code--defined as $-1$ eigenstates of the $A_v$ and $B_p$ operators, located on the vertices and plaquettes of the square lattice, respectively.  Because of this, an additional error pathway exists in the protocol where defects of different type are uncontrollably wound around one another.  This can be suppressed by operating at low temperature.  Secondly, both of these defects undergo two-dimensional random walks rather than one-dimensional random walks.  This difference in dimension gives rise to a modified form of the toric code's finite temperature error rate, due to the differing nontrivial topological random walk probability for two dimensions versus one.

Operationally, these differences only require small modifications of the autonomous protocol.  Namely, there need be two $\rm{DSWAP}$ operators:

\begin{align}
\rm{DSWAP}^e_{vv'}&=\frac{1}{4}{\sigma }^x_{vv'}\left(1-A_{v}\right)\left(1+A_{v'}\right) \\ \label{eq:TCDSWAPs}
\rm{DSWAP}^m_{pp'}&=\frac{1}{4}{\sigma }^z_{pp'}\left(1-B_{p}\right)\left(1+B_{p'}\right)
\end{align}

These translate an $A$-type ($B$-type) excitation from a vertex $v$ (plaquette $p$) to an adjacent vertex $v'$ (plaquette $p'$).  Second, the $\lambda$-mixing protocol shuttles defects towards a shared boundary of length $\lambda$ between subdomains of charactersitic area $\lambda^2$.

Because subregions share a boundary of length $\lambda$ rather than a single site, as in the one-dimensional case, the cycling protocols require at most a factor of $\lambda$ more swaps to complete a cycle.  The protocol then takes the following simple form:

1. Choose a species of quasiparticle

2. Divide the lattice into domains of characteristic area $\lambda^2$

3. Pick two $\lambda$-domains which share a boundary

4. Pick two defect locations within these two $\lambda$-domains.

5. If these defect locations are within the same $\lambda$-domain, apply $\rm{DSWAP}$s until they would be nearest neighbors.  If they are in different $\lambda$-domains, apply $\rm{DSWAP}$s until they meet at the shared boundary.

6. Repeat (5) until all pairs of defect locations are exhausted.

7. Repeat (3) through (6) until all pairs of $\lambda$ domains which share a boundary are exhausted.

8. Repeat (1) through (7) until all species of quasiparticle are exhausted.

This protocol is also highly parallelizable, both by operating on multiple pairs of $\lambda$-domains, and by acting on simultaneous pairs of defect sites within pairs of $\lambda$-domains.

\subsection{The General Problem}

We can always divide a $d$-dimensional lattice into $N \equiv L^d/\lambda^d$ domains and try to devise an algorithm that fuses defects between adjacent domains. For our protocol, defects are shuttled towards $d-1$-dimensional boundaries between adjacent domains of volume $\lambda^d$.  From this, we can generalize the low temperature dynamics of equation \eqref{eq:lowchidynamics} to the $d$-dimensional case as follows:

\begin{equation}
\Gamma_{\rm{Cyc}} \propto L^d \gamma_+ \frac {\gamma_0}{\lambda \chi} \frac{\gamma_0}{\gamma_-} P_{\Omega}^d(L,\lambda) \frac{f(\lambda^d)}{L^d}, \label{eq:ddimdynamics}
\end{equation}

where $L$ is the edge length of the $d$-dimensional volume enclosed by the system, $P_{\Omega}^d(L,\lambda)$ encodes the probability of a nontrivial topological random walk of a pair of defects in $d$ dimensions with system size $L$ and domain lengthscale $\lambda$, and $f(\lambda^d) \propto \lambda^d$ is a protocol dependent function which depends on the implementation details of the algorithm.

Theorem \eqref{th:polyscale} guarantees that an algorithm exists which can perform the cycling in a number of steps polynomial in the dimension of the lattice.  However, this Theorem does not guarantee that a $\lambda$-mixing protocol exists which solves the problem.  In general, for higher dimensions, there are always defect patterns of distance $O(\lambda)$ which are uncorrectable by our $\lambda$-mixing protocol.  The design strategy is then to try and maximize this minimum uncorrectable distance by careful tiling of the graph of interest.

\begin{figure}
\begin{center}
\scalebox{1}{\includegraphics[width=1.0\columnwidth]{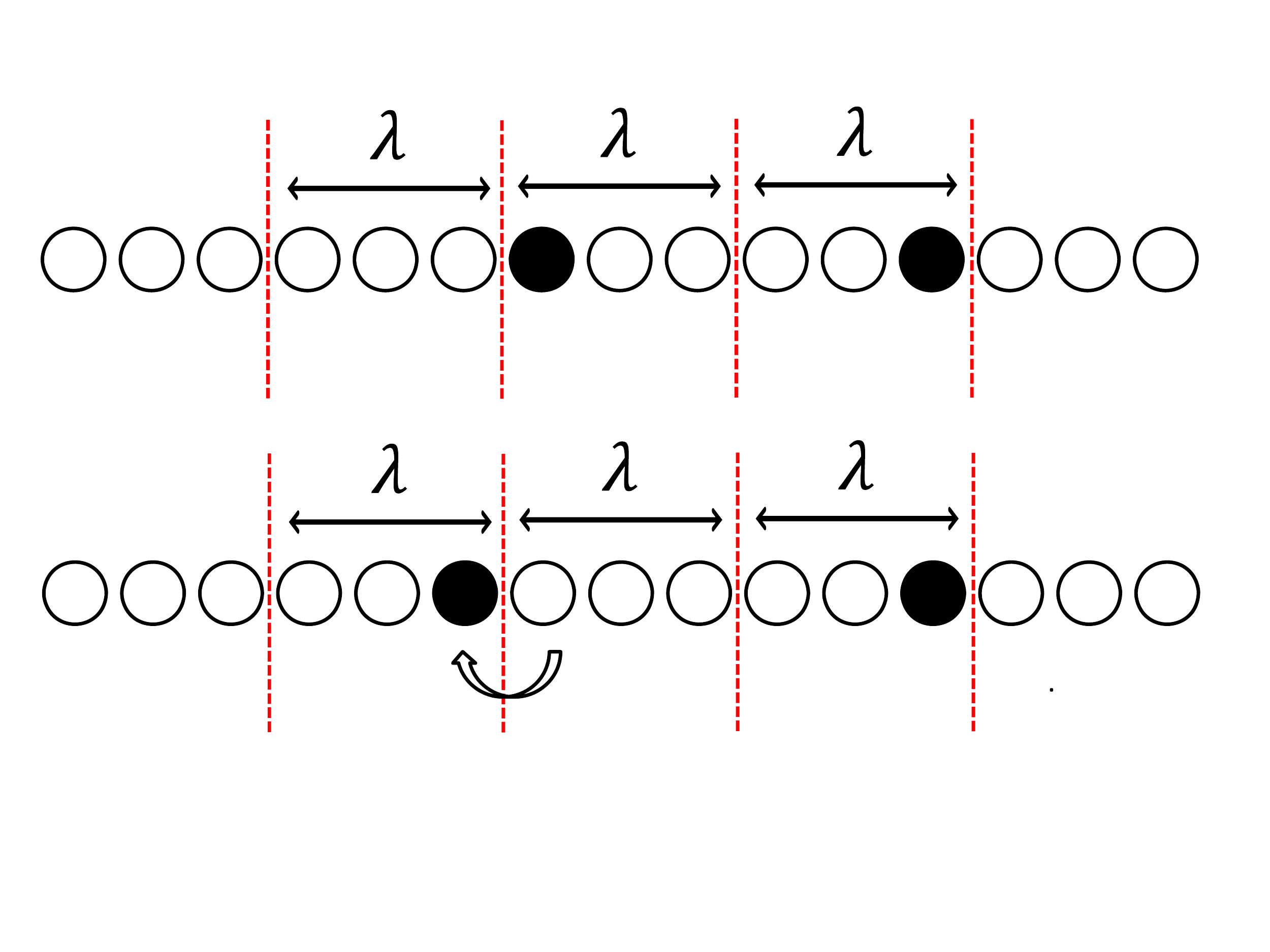}}
\end{center}
\caption{A single random application of the bath hopping operator causes a defect to move between adjacent regions of size $\lambda$.  Once a pair of defects are separated by this distance, the protocol will not be able to correct them with certainty.}
\label{fig:defectescape}
\end{figure}

To be more explicit, if a single pair of defects appears on the graph, uncorrectable errors are generated only when one of the defects escapes to an adjoining region which does not share a boundary with its pair.  A cartoon of this process is depicted in figure \figref{fig:defectescape}.
For one dimension, this cannot happen in one step after an adjacent pair of defects appears.  More specifically, if a single pair of adjacent defects appears on the lattice, no single DSWAP will cause such an error to occur, by design, and no single bath operation will cause an adjacent pair of defects to be in nonadjacent regions of size $\lambda$.  In one dimension, at worst a pair will be created, shuttled around by the protocol, and \emph{then} translate by a bath operator across a boundary, resulting in an uncorrectable error.

\begin{figure}
\begin{center}
\scalebox{1}{\includegraphics[width=1.0\columnwidth]{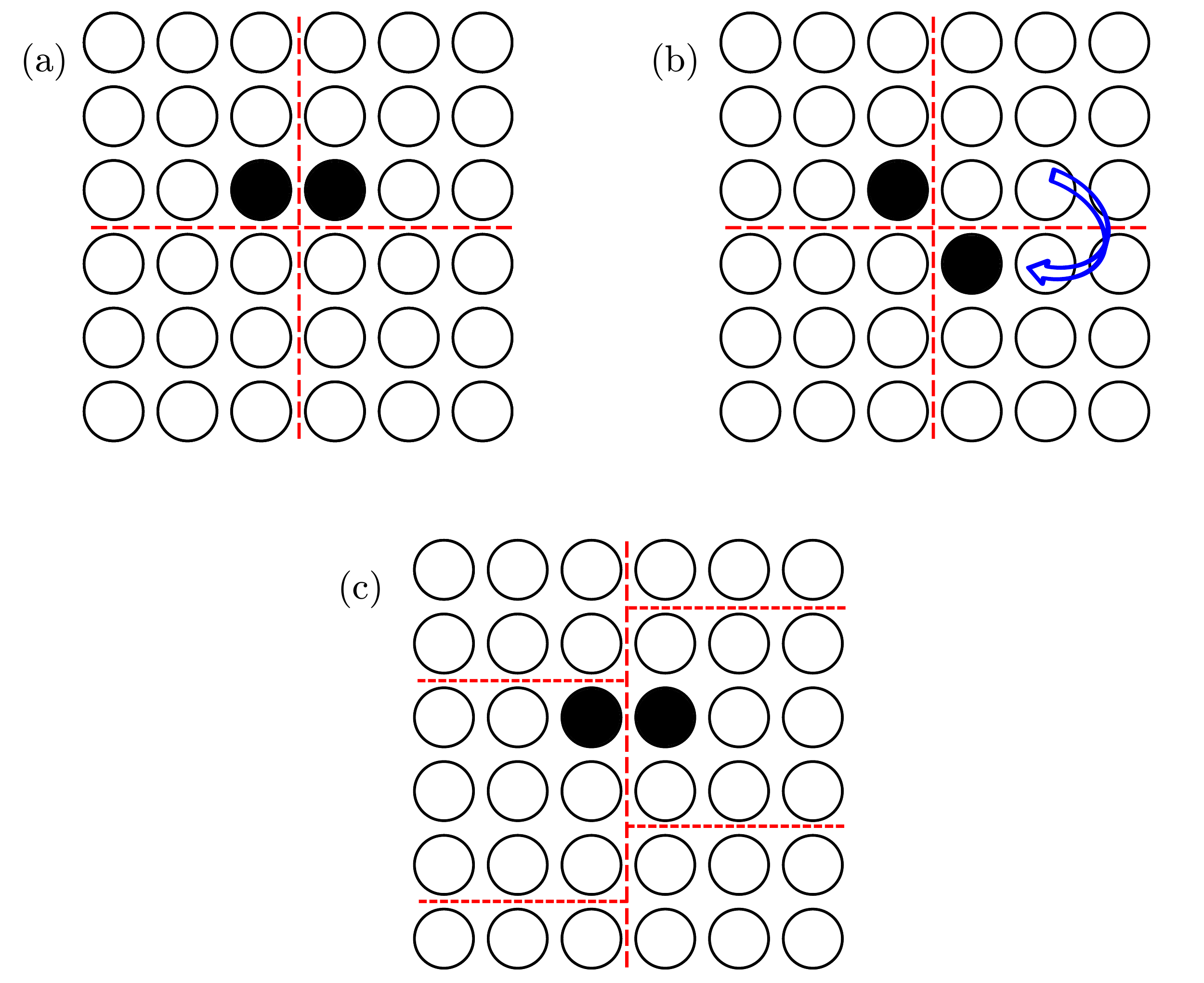}}
\end{center}
\caption{A single random application of the bath hopping operator causes a defect to move between nonintersecting regions of size $\lambda$, depicted in (a) and (b).  (c) depicts a new lattice tiling where no single swap can move defects into two $\lambda$ regions which do not share a boundary.}
\label{fig:badgooddesign}
\end{figure}

This distinction is important because poor choice of tiling in higher dimension \emph{can} result in uncorrectable errors that occur in a single step after pair creation.  For example, compare the single hop in the upper half of \figref{fig:badgooddesign} to the lower half.  A defect pair appearing at a corner can transition to an uncorrectable configuration in a single step, whereas in the lower tiling, this is not possible for any initial configuration of adjacent defect pairs.  This can be checked by simple enumeration of the possible defect locations and single-hop geometries.

This shifted square lattice tiling depicted in the lower half of Figure \ref{fig:badgooddesign} generalizes to three dimensions, and is necessary for equation \eqref{eq:ddimdynamics} to describe the leading order error process.

\subsection{Hybrid \rm{DSWAP}-Stabilizer Codes}
\label{sec:hybridcodes}

While we have demonstrated that our protocol gives rise to an enhanced lifetime for a topological code with string-like error operators, it is also possible and desirable, to use our $\rm{DSWAP}$ cycling protocol with a more traditional stabilizer detection-correction scheme simultaneously.  We postpone numerical analysis of such a scheme for future work, but we sketch such a protocol in this section.

For concreteness, we specialize here to the case of the toric code.  Error detection and correction in the toric code requires (1) measurement of all stabilizer syndrome operators and (2) application of a perfect-matching algorithm to determine which pairs of defects to fuse.  Whether or not such an algorithm will be successful depends on the density of defects at the time of measurement.  Given a stabilizer measurement rate $\gamma$, in the ``infinite temperature'' limit, each site is found to have a defect with probability $p$, independently, except for the ensemble requirement that the total number of defects be even.  It is well known that if $p$ is below some critical value, $p_c$, it is possible to correct the errors in the toric code with certainty. Equivalently, $p_c$ sets the minimum rate at which measurement must occur so that detection is possible in principle.  Call this rate $\gamma_c$.

For the protocol to have an effect, we must operate in a regime where pair annihilation is favored over pair creation.  For simplicity, we work in the low temperature regime where single defect pairs dominate. In this regime, an uncorrectable error has occurred when a single pair of defects becomes separated by more than half the linear lattice dimension.  In the presence of the \rm{DSWAP} cycling protocol, the rate associated with such an event occuring is modified by some constant factor:

\begin{equation}
\frac{1}{\Gamma_{\rm{Toric\ Code\ Cycling}}} = g \frac{1}{\Gamma_{\rm{Toric\ Code}}} \label{eq:TCenhancement}
\end{equation}

where $\Gamma_{\rm{Toric\ Code\ Cycling}}$ is the error rate of the toric code in the presence of a cycling protocol, and $\Gamma_{\rm{Toric\ Code}}$ is the error rate in the absence of the protocol.

For $g > 1$, i.e., when our protocol actually enhances the lifetime of the code, this effectively reduces the critical detection rate $\gamma_c$ by the same factor.  This is because the protocol effectively reduces the rate at which undetectable pairs are created.

Thus, if a physical realization of a stabilizer error detection/correction cycle is rate limited due to hardware or fundamental noise constraints, the $\rm{DSWAP}$ cycling protocol provides one avenue towards reducing the critical measurement/detection rate purely by application of local unitaries.

\section{Discussion}
\label{sec:Discussion}

We have provided a dissipative error correction protocol that enhances the lifetime for models with string-like defects.  In particular, we have derived an enhanced lifetime for the one-dimensional Ising model in the presence of our protocol, i.e. equation \eqref{eq:lowchidynamics}, and provided numerical evidence for this enhancement given certain rate assumptions.  Practically, this algorithm increases the lifetime of the system linearly with system size up to a system-size independent cutoff, as anticipated from No-Go theorems.  Furthermore, we have sketched how this protocol can be generalized to higher dimensional models like the toric code, and used in conjunction with traditional stabilizer error detection/correction schemes.

The efficacy of these sorts of protocols is intimately related to the scaling of the protocol with system size and protocol parameters, as we have demonstrated.  Notably, the best performing versions of our protocol have small $\lambda$, and, in a sense, only correct the shortest distance errors.  This may seem counterintuitive from the perspective of designing protocols which correct as many errors as possible.  For example, suppose we wish to compare a $\lambda=3$ protocol with total cycle time $\tau$ to a $\lambda=4$ protocol with the same cycle time $\tau$.  Note that by fixing total cycle time, we are implicitly requiring that the $\lambda=4$ protocol be performed more quickly at the level of individual application of $\rm{DSWAP}$s (because there are more $\rm{DSWAP}$s in a complete cycle), but we require that the complete error correcting cycle of each protocol is completed in the same amount of time.  Naively, we would expect the $\lambda=4$ protocol to do better, because it is dissipating errors over a longer length scale, but in the same amount of time.  \figref{fig:serialmemscaling} indicates a narrow region where this is the case, but, generically, this is not the case.

This can be traced to the poor scaling of the maximal lifetime with $\lambda$, as represented by equation \eqref{eq:chicrit}.  Essentially, the protocols which correct larger distance errors---i.e., large $\lambda$-fixing protocols---employ so many gates that all of the gains of correcting longer distance errors are erased by the time it takes to actually perform the protocol, even when implementing the protocol in parallel.

Additionally, the form of equation \eqref{eq:lowchidynamics} suggests that larger systems manifestly have lower error rates, because $\Gamma_{\rm{Cyc}} \propto \frac{1}{\chi \L}$ for low temperature and $L >> \lambda$.  For sufficiently large systems this once again breaks down due to equation \eqref{eq:chicrit}.  Namely, for a fixed cycling rate $\chi_0$, making the system larger only increases the lifetime so long as $\chi_0 < \chi_c$.

These shortfalls could be circumvented by allowing for longer range unitaries.  For example, the $\rm{DSWAP}$ operator could be replaced by a generalized operator $\rm{DSWAP}_\lambda$ which transports domain walls over longer distances.  Our insistence on building the protocol entirely out of local $\rm{DSWAP}$ gates was to perform as honest an analysis as possible with respect to the power of this type of protocol.  But if a particular architecture could exchange defects over long distances just as easily as short ones, this would immediately allow for algorithms with better scaling.  We hope to examine the optimality of these sorts of protocols in future work.

In the long term, this program is meant to identify the simplest possible set of ingredients necessary to provide protection for a stabilizer code based quantum memory.  Many partial ingredients are known, like the No-Go theorems mentioned in Sec. \ref{sec:Intro}.  Practically, the goal is a protocol designed around the dynamics of the excitations of the stabilizer codes of interest with miminimal usage of resources, but which still results in an error threshold so that a state can be preserved indefinitely.  With this work, we have demonstrated a constant factor improvement with only local unitaries dressing the system.

\section{Acknowledgments}
This material is based upon work supported by DARPA under Grant No. 3854-UCB-AFOSR-0041 and by the National Science Foundation under Grants No. PIF-0803429 and No. CHE-1213141.  CDF was supported by the NSF Graduate Research Fellowship under Grant DGE-1106400.

\appendix
\section{Syndrome Decoding for 1D Ising Model}
\label{sec:syndecode}

The corrective operator $O$ in \figref{fig:lambda3ECC} can be written as a collection of conditional applications of $\rm{DSWAP}$ and $\rm{DWALL}$, where the applications of the operators are conditioned on the measurements of the stabilizers.  We adopt the notation $\rm{O}_{123}$ to indicate the application of the operator $O$ on qubits $1, 2$ and $3$.  Then, in table form, the operator $O$ is:

\begin{table}
\begin{center}
  \begin{tabular}{ c | c | c | c}
    \hline
    $s_1$ & $s_2$ & $s_3$ & $O$  \\ \hline
    $1$ & $1$ & $1$ & $I$  \\ \hline
    $1$ & $1$ & $-1$ & $I$  \\ \hline
    $1$ & $-1$ & $1$ & $I$  \\ \hline
    $1$ & $-1$ & $-1$ & $\rm{DWALL}_{234}$  \\ \hline
    $-1$ & $1$ & $1$ & $I$  \\ \hline
    $-1$ & $1$ & $-1$ & $\rm{DSWAP}_{123} \rm{DWALL}_{234}$  \\ \hline
    $-1$ & $-1$ & $1$ & $\rm{DWALL}_{123}$  \\ \hline
    $-1$ & $-1$ & $-1$ & $I$  \\
    \hline
  \end{tabular}
\end{center}
\caption{Corrective operations given certain measurements of the stabilizers $s_1$ through $s_3$ in \figref{fig:lambda3ECC}.  If an odd number of domain walls are detected, the identity is applied.}
\end{table}

\section{\rm{MATCHSEQ} and Error Correction}
\label{sec:MATCHSEQ}

\subsection{Polynomial Scaling}
Define the game $MATCHSEQ$ as follows:  two nonadjacent vertices on a simply connected graph $G$ are colored black, called defects, the rest white.  The player is allowed to perform a conditional swap, or DSWAP, on any two adjacent vertices, which exchanges black vertices and white vertices, and does nothing to pairs of white vertices.  If black vertices become adjacent, they immediate \emph{fuse} and become white vertices.  Crucially, the player does not know which vertices are colored black.

``Winning'' $MATCHSEQ$ amounts to performing a sequence of moves which guarantees that a pair of arbitrarily placed vertices fuses.

Define the pairing sequence $M(G_{v})$ to be the sequence of conditional swaps necessary to bring any configurations of two defects adjacent to one another at least once on a graph $G$ with $v$ vertices.  Define the pairing number $|M(G_v)|$ to be the pairing sequence with minimal length.  Table 1 tabulates the first few nontrivial pairing numbers for the special case of G equal to a linear chain of length L.

\begin{table}
\begin{center}
  \begin{tabular}{ l | r }
    \hline
    $|M(3)|$ & 1 \\ \hline
    $|M(4)|$ & 3 \\ \hline
    $|M(5)|$ & 6 \\ \hline
	$|M(6)|$ & 10 \\ \hline
	$|M(7)|$ & 18 \\
    \hline
  \end{tabular}
\end{center}
\caption{Minimum number of DSWAPs required to necessarily fuse any two defects on a linear chain with open boundary conditions.  Computed via breadth first search.}
\end{table}

\begin{theorem}
The number of DSWAPs necessary to win $MATCHSEQ$ for an arbitrary finite, connected graph $G$ is polynomial in the number of vertices in the graph $G$. \label{th:polyscale}
\end{theorem}

Proof: Let $M^{*}(G)$ be a winning strategy on an arbitrary graph $G$.  Suppose an arbitrary vertex is added to $G$, called $v^*$, with up to $|G|$ edges.  Call this modified graph $G'$.  Then, performing $M^{*}(G)$ on $G'$ either fuses two arbitrarily placed defects, or there's a single defect on the new vertex, and the remaining vertex has just been permuted around in $G$.   A candidate $M(G')$ is then: \newline
1.	Perform $M^{*}$ on $G$. \newline
2.	Pick a vertex, $v'$, on $G$.  Supposing a defect is on $v'$, perform a sequence of DSWAPs that brings that defect adjacent to the new vertex, $v^*$. \newline
3.	Perform the reverse of the sequence of DSWAPs in (2), and repeat (2) with a new $v'$. \newline
4.	Repeat (2) and (3) until all vertices in G are exhausted. \newline

The number of DSWAPs needed for step (2) is at most $|v|$, i.e., the number of vertices in $G$.  Thus, the total complexity of steps 2 through 4 is $O(|v|^2)$.  This admits a recurrence relation:

\begin{equation}
|M(G_{v+1})| \leq |M(G_v)| + b*|v|^2,
\end{equation}

Where $b$ is a constant $\leq 1$.  $b=1$ corresponds to the case that the new vertex is only connected to one vertex in the original graph $G$.  Solving this recurrence relation in the limit that the inequality is always saturated yields $|M(G_v)| \leq O(|v|^3)$.  It is worth emphasizing that this is not the minimal such solution to MATCHSEQ, just one that is easily provably polynomial in $|G|$.  The likely graph structures of interest to an experimentalist, i.e., linear chains, square lattices, admit more favorable algorithms with softer polynomial scaling.

\subsection{Strategies}

For a given winning strategy, $M^{*}(G)$, it will be convenient to classify the strategy based on the maximum distance that any given defect is moved.  In the sequel, we will construct $M^{*}(G)$ out of a concatenation of $M^{*}(G_i)$, where $G_i$ are subgraphs of $G$.  Thus, if only a single defect happens to be in the subgraph $G_i$, we would like to bound the maximum displacement of that defect by the strategy.

Let $d(\emph{M})$ be the maximum distance any given defect is moved by a given strategy.  For winning strategies, $d(\emph{M})$ is at least half of the maximum distance between defects and at most permutes defects around the entire graph, so $|v| \geq d(\emph{M}) \geq |v|/2$.  Define a strategy $M^{*}(G)$ to be \emph{$k$-mixing} if $d(\emph{M}) = k$.

We introduce this terminology because most physical realizations of \rm{MATCHSEQ} will have a background rate of uncontrollable \rm{DSWAP}s, driven by coupling to a bath.  $k$-mixing strategies are necessary in such cases to be partially resilient to these random ``error'' \rm{DSWAP}s.

\begin{figure}
\begin{center}
\scalebox{1}{\includegraphics[width=1.0\columnwidth]{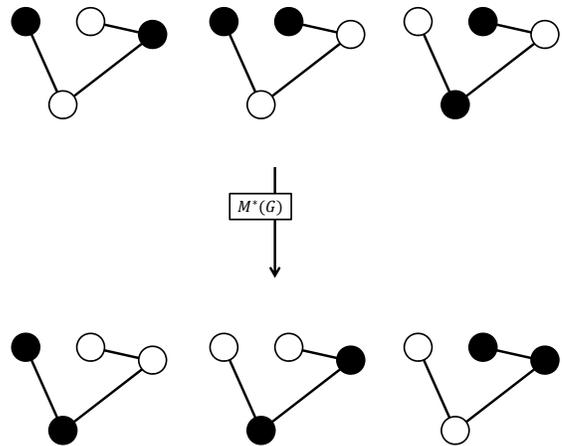}}
\end{center}
\caption{An example of the result of applying a winning sequence $M(G)$ to a graph with defect pairs present.  No matter where the defects are, the sequence of DSWAPs brings pairs adjacent, whereupon they immediately fuse.}
\label{fig:MATCHSEQexample}
\end{figure}

\subsection{Mapping onto 1D Ising Model}

Vertices in the problem setup for \rm{MATCHSEQ} correspond to the dual lattice of the Ising chain, and the process of fusion is simply dissipation of adjacent domain wall pairs by the bath.

However, we caution that the mapping onto \rm{MATCHSEQ} is only partial: defects on the Ising chain hop in the absence of any experimental intervention, so the Ising chain is more akin to a game of MATCHSEQ with a random, background DSWAP rate.  Further, there can be more than two pair of excitations on the Ising chain, but for low temperature, the regime where the protocol works best, this is exceedingly rare.  Lastly, defect pairs don't necessarily fuse immediately--fusion happens at the timescale set by the system-bath coupling, the type of bath model, and the temperature, so rate at which \rm{DSWAP}s are applied must be chosen carefully for optimal lifetime enhancement.

\section{Algorithm for 1-D Ising Model}
\label{sec:Algorithm}

Here we provide python code for a $\lambda$-mixing algorithm for the Ising chain.  The output of the algorithm is a sequence of locations.  Our convention is such that location $i$ indicates a \rm{DSWAP} should be applied that exchanges defects between sites $i$ and $i+1$.   Heuristically, the algorithm attempts to shuffle defects towards the shared boundary of the disjoint sites {0,1,2,...,$\lambda$-1} and {$\lambda$,$\lambda$+1,...,2$\lambda$-1}.  That is, it attempts to translate defects so that they are adjacent to each other at sites $\lambda-1$ and $\lambda$.  After completing this cycle, the algorithm repeats for the next two adjacent domains, $\lambda,...,2\lambda-1$ and $2\lambda,...,3\lambda-1$.  This continues until the lattice has been exhausted.

The following python code generates a complete sequence of \rm{DSWAP}s given a lattice size and $\lambda$ length.

\begin{lstlisting}
def SwapProtocol(L, lamb):
	prot = []
	numofdomains = L / lamb
	for d in xrange(numofdomains):
		for k in xrange(lamb):
			for m in xrange(k):
				prot.append((lamb-1-k+m+d*lamb)%L)
			for i in xrange(lamb):
					for j in xrange(i):
						prot.append((lamb+i-j-1+d*lamb)%L)
    return prot
\end{lstlisting}

The protocol is parallelized by operating simultaneously on specific pairs of domains.  To be more explicit: denote the first $\lambda$ sites as $\lambda_1$, the next $\lambda$ sites $\lambda_2$ and so on.  The algorithm can be naturally partitioned into a sequence of \rm{DSWAP}s that translates defects to the shared boundary of $\lambda_1$ and $\lambda_2$ (call this sequence ($\lambda_1,\lambda_2$)), followed by a sequence that translates defects to the shared boundary between $\lambda_2$ and $\lambda_3$, (call this sequence ($\lambda_2,\lambda_3$)), etc.  To parallelize, apply the sequence ($\lambda_1,\lambda_2$) simultaneously with ($\lambda_3,\lambda_4$), ($\lambda_5,\lambda_6$), etc.  When complete, apply the sequence ($\lambda_2,\lambda_3$) with ($\lambda_4,\lambda_5$) etc.  This exhausts the protocol.

\bibliography{manuscript_TC_2}

\end{document}